\documentclass[a4paper]{article}
\pdfoutput=1

\usepackage{graphicx,wrapfig}
\usepackage{lipsum}
\usepackage{amsmath,amssymb,mathrsfs}
\usepackage{hyperref}
\usepackage{url}
\usepackage{mathtools}
\usepackage{changepage}
\usepackage{multirow}
\usepackage{wrapfig}
\usepackage{subfig}
\usepackage{algorithm}
\usepackage[noend]{algpseudocode}
\usepackage{color}
\usepackage{epsfig,enumerate,amsmath,amsfonts,amssymb,amsthm,mathrsfs,ifpdf}
\usepackage{indentfirst,relsize}
\usepackage[numbers]{natbib}
\usepackage{setspace,graphicx}
\usepackage{latexsym}
\usepackage[all]{xy}
\usepackage[usenames,dvipsnames]{pstricks}
\usepackage{pst-grad} 
\usepackage{pst-plot} 

\usepackage[margin = 2.50cm]{geometry}

\newcommand{\remove}[1]{}

\newtheorem{theo}{Theorem}
\newtheorem{lem}[theo]{Lemma}

\newtheorem{cl}[theo]{Claim}
\newcounter{Ca}[theo]
\newtheorem{ca}[Ca]{Case}
\newtheorem{defi}[theo]{Definition}
\newtheorem{remk}{Remark}

\begin{document}

\title{On vertex-edge and independent vertex-edge domination}

\author{Subhabrata Paul\footnote{Department of Mathematics, IIT Patna, India} \and Keshav Ranjan\footnote{Department of Computer Science \& Engineering, IIT Madras, India} \thanks{This work was done when the second author was persuing his M.Tech. at IIT Patna.}}

\maketitle              

\begin{abstract}
Given a graph $G = (V,E)$, a vertex $u \in V$ \emph{ve-dominates} all edges incident to any vertex of $N_G[u]$. A set $S \subseteq V$ is a \emph{ve-dominating set} if for all edges $e\in E$, there exists a vertex $u \in S$ such that $u$ ve-dominates $e$. Lewis [Ph.D. thesis, 2007] proposed a linear time algorithm for ve-domination problem for trees. In this paper, first we have constructed an example where the proposed algorithm fails. Then we have proposed a linear time algorithm for ve-domination problem in block graphs, which is a superclass of trees. We have also proved that finding minimum ve-dominating set is NP-complete for undirected path graphs. Finally, we have characterized the trees with equal ve-domination and independent ve-domination number.
\end{abstract}

\section{Introduction}
\label{sec:intro}

Domination and its variants are one of the classical problems in graph theory. Let $G=(V,E)$ be a graph and $N_G(v)$ (or $N_G[v]$) be the \emph{open} (respectively, \emph{closed}) neighborhood of $v$ in $G$. A set $D\subseteq V$ is called a \emph{dominating set} of a graph $G=(V,E)$ if $|N_G[v] \cap D| \geq 1$ for all $v \in V$. Our goal is to find a dominating set of minimum cardinality which is known as \emph{domination number} of $G$ and denoted by $\gamma(G)$. For details the readers are refered to \cite{HaynesHedetniemiSlater11998,HaynesHedetniemiSlater1998}.

In this paper, we have studied one variant of domination problem, namely \emph{vertex-edge domination} problem, also known as \emph{ve-domination} problem. Given a graph $G = (V,E)$, a vertex $u \in V$ \emph{ve-dominates} all edges incident to any vertex of $N_G[u]$. A set $S \subseteq V$ is a \emph{vertex-edge dominating set} (or simply a \emph{ve-dominating set}) if for all edges $e\in E$, there exists a vertex $u \in S$ such that $u$ ve-dominates $e$. The minimum cardinality among all the ve-dominating sets of $G$ is called the \emph{vertex-edge domination number} (or simply \emph{ve-domination number}), and is denoted by $\gamma_{ve}(G)$. A set $S$ is called an \emph{independent ve-dominating set} if $S$ is both an independent set and a ve-dominating set. The \emph{independent ve-domination number} of a graph $G$ is the minimum cardinality of an independent ve-dominating set and is denoted by $i_{ve}(G)$.

The vertex-edge domination problem was introduced by Peters\cite{Peters1986} in his PhD thesis in $1986$. However, it did not receive much attention until Lewis\cite{lewis2007vertex} in $2007$ introduced some new parameters related to it and established many new results in his PhD thesis. In his PhD thesis, Lewis has given some lower bound on $\gamma_{ve}(G)$ for different graph class like connected graphs, $k$-regular graphs, cubic graphs etc. On the algorithmic side, Lewis has also proved that the ve-domination problem is NP-Complete for bipartite, chordal, planar and circle graphs and independent ve-domination problem is NP-Complete even when restricted to bipartite and chordal graph. Also approximation algorithm and approximation hardness results are proved in \cite{lewis2007vertex}. In \cite{LewisHedetniemiHaynesFricke2010}, the authors have characterized the trees with equal domination and vertex-edge domination number. In \cite{KRISHNAKUMARI2014363}, both upper and lower bounds on ve-domination number of a tree have been proved. Some upper bounds on $\gamma_{ve}(G)$ and $i_{ve}(G)$ and some relationship between ve-domination number and other domination parameters have been proved in \cite{Boutrig2016}. In \cite{Zylinski2019}, {\.{Z}}yli{\'{n}}ski has shown that for any connected graph $G$ with $n\geq 6$, $\gamma_{ve}(G)\leq n/3$. Other variations of ve-dominations have also been studied in literature \cite{BoutringTotalVE2018,Krishnakumari2017}.

\begin{wrapfigure}{r}{0.450\textwidth}
	\begin{center}
		\includegraphics[width=5cm]{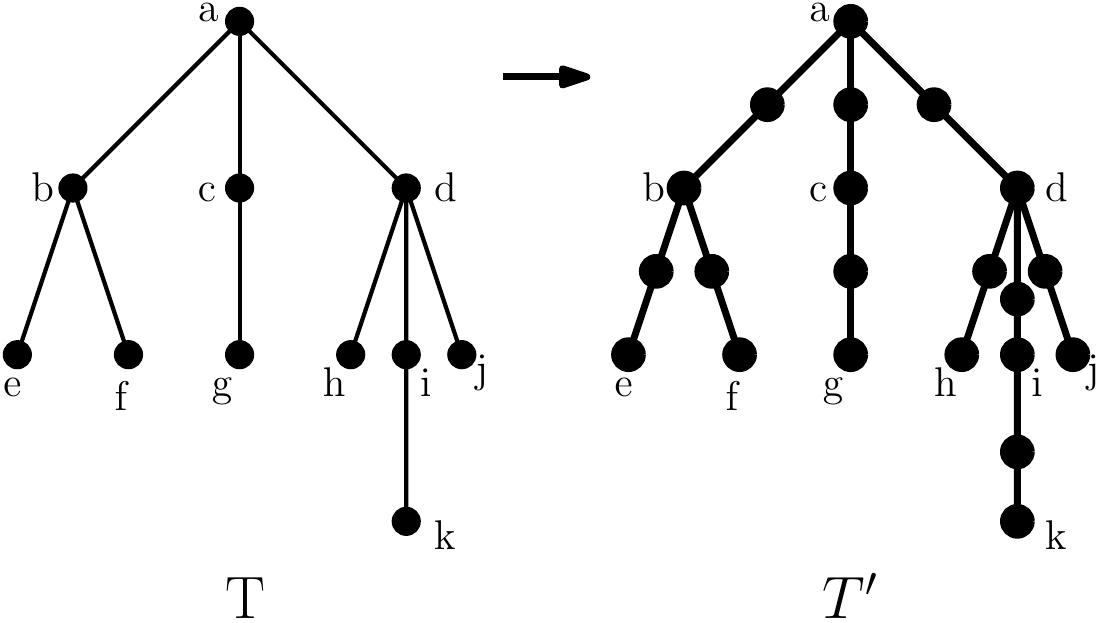}
		\caption{Counter example}
		\label{fig:cntr_exmpl}
	\end{center}
\end{wrapfigure}  
In \cite{lewis2007vertex}, Lewis proposed a linear time algorithm for ve-domination problem for trees. Basically, he proposed a linear time algorithm for finding minimum distance-$3$ dominating set of a weighted tree. A set $D\subseteq V$ is called a \emph{distance-$3$ dominating set} of a graph $G=(V,E)$ if every vertex in $V$ is at most distance $3$ from some vertex in $D$. In case of a weighted graph, the goal is to find \emph{distance-3 dominating set} with minimum weight. Lewis claimed that ``Given any tree $T=(V,E)$, define a new tree $T'=(V',E')$ by subdividing each edge of $E$. Then place a weight of  one on each of $V$ and a weight of $\infty$ on each of $V'\setminus V$. Now solve the weighted distance-$3$ dominating set problem for $T'$, the result will give the $\gamma_{ve}(T)$." But, we have found a counter example of this claim. In Figure \ref{fig:cntr_exmpl},  it is easy to see that $\gamma_{ve}(T) =2$. Now, in the new weighted tree $T'=(V',E')$, it is not possible to find any distance-3 dominating set whose weight is $2$.

This motivates us to study ve-domination problem in trees and other graph classes. The rest of the paper is organized as follows. In Section \ref{sec:Block_graph}, we have proposed a linear time algorithm for finding minimum ve-dominating set in block graphs, which is a superclass of trees. Section \ref{sec:UDPgraph} deals with NP-completeness of this problem in undirected path graphs. In Section \ref{sec:VEequalIVE}, we have characterized the trees having equal ve-domination number and independent ve-domination number. Finally Section \ref{sec:conclu} concludes the paper. 
\section{Block graph}
\label{sec:Block_graph}
In this section, we propose a linear time algorithm for block graphs to solve ve-domination problem. A vertex $v\in V$ is called a \emph{cut vertex} of $G=(V,E)$ if removal of $v$ increases the number of components in $G$. A maximal connected induced subgraph without a cut vertex of $G$ is called a \emph{block} of $G$. A graph $G$ is called a \emph{block graph} if
each block of $G$ is a complete subgraph. The intersection of two distinct blocks can contain at most one vertex. Two blocks are called \emph{adjacent blocks} if they contain a common cut vertex of $G$. A block graph with one or more cut vertices contains at least two blocks, each of which contains exactly one cut vertex. Such blocks are called \emph{end blocks}. The \emph{distance between two blocks} $B_i$ and $B_j$ is defined as $dist_G(B_i,B_j)= max\{dist(v_i,v_j)|v_i\in B_i, v_j\in B_j\} - 1$. The \emph{distance between a vertex $v$ and a block $B$ } of a block graph $G$ is denoted as $dist_G(v,B)= max\{dist_G(v,u)|u \in B\} - 1$.

Our proposed algorithm is a labeling based algorithm. Let $G=(V,E)$ be a block graph where each vertex $v\in V$ is associated with a label $l(v)$ and each edge $e=(xy)\in E$ is associated with a label $m(xy)$, where $l(v)$, $m(xy)\in \{0,1\}$. We call such graph a labelled block graph. 
\begin{defi}
Given a labelled block graph $G=(V,E)$ with labels $l$ and $m$, we first define an \emph{optional ve-dominating set} as a subset $D\subseteq V$ such that 
\begin{enumerate}
\item[$(i)$] if $l(v)=1$, then $v\in D$,
\item[$(ii)$] $D$ ve-dominates every edge $e=(xy)$ with $m(xy)=1$.
\end{enumerate}

The \emph{optional ve-domination number}, denoted by $\gamma_{opve}(G)$, is the minimum cardinality among all the optional ve-dominating sets of $G$.
\end{defi}

Note if $l(v)=0$ for all $v\in V$ and $m(xy)=1$ for all $e=(xy)\in E$, then the minimum optional ve-dominating set is nothing but a minimum ve-dominating set of $G$. Given a labelled block graph $G$ with labels $l(v)=0$ for each $v\in V$ and $m(xy)=1$ for each $xy\in E$, our proposed algorithm basically outputs a minimum optional ve-dominating set of $G$. 

Next we present the outline of the algorithm. Let $B_0$ be an end block of a block graph $G=(V,E)$. Since, block graph has a tree like structures, we can view $G$ as a graph rooted at the end block $B_0$. The \emph{height} of $G$ is defined as $h(G)=\max\{dist_G(B_0,B)|\text{B is end block of G}\}$. At each step, the algorithm processes one of the end blocks at maximum height. Moreover, out of all the blocks at the same maximum height, the blocks with more number of edges with $m(xy)=1$, are processed first. Based on some properties of the edges having label $1$ in an end block $B$, we decide whether to take some vertices from $B$ in the optional ve-dominating set or not and then delete that block $B$ (except the cut vertex) from $G$. We also modify the labels of some of the vertices and edges of the new graph. In the next iteration, we process another end block and the process continues till we are left with the root block $B_0$. For the root block, we directly calculate the optional ve dominating set. The outline of the algorithm is given in Algorithm \ref{alg:minVEDB}. In Algorithm \ref{alg:minVEDB}, for an end block $B$, $t_B$ denotes the number of edges with $m(xy)=1$ in $B$ and $P$ denotes the set of non-cut vertices of $B$, i.e., $P=V(B)\setminus \{c\}$, where $V(B)$ is the set of vertices of $B$ and $c$ is the cut vertex of $B$. Also let $F(c)$ denote the unique cut vertex of $G$ in $N_{G}[c]$ which has the minimum distance from $B_0$. Note that $F(c)=c$ if and only if $c$ is the cut vertex of $B_0$.

\begin{algorithm}[h!]
	\caption{MIN\_OPT\_VEDom(G)} 
	\label{alg:minVEDB}
	\textbf{Input:}A labelled block graph $G=(V,E)$ with $l(v)=0,\forall v\in V$ and $m(xy)=1,\forall xy\in E$\\
	\textbf{Output:} Minimum optional ve-dominating set $S$ of $G$
	\begin{algorithmic}[1]
		\State $S=\phi$
		\For{\texttt{i=$h(G)$ to 1}}
		\State $B$ is an end block at level $i$ and $c$ is the cut vertex of $B$
		\While{($t_B \geq 2$)}
		\State $l(c)=1$
		\State $G=G \setminus P$
		\State $m(xy)=0$, $\forall x \in N_{G}(c)$
		\EndWhile
		\While{(($t_B = 1$ such that $m(uv)=1$) \texttt{AND} $(u \not = c)$ \texttt{AND} $(v\not = c)$)}
		\State $l(c)=1$
		\State $G=G \setminus P$
		\State $m(xy)=0$, $\forall x \in N_{G}(c)$
		\EndWhile
		\While{($t_B = 1$)}
		\State $l(F(c))=1$
		\State $G=G \setminus P$
		\State $m(xy)=0$, $\forall x \in N_{G}(F(c))$
		\EndWhile
		\While{($t_B = 0$)}
		\State $S= S \cup \{x | x \in P \text{ and } l(x)=1\}$
		\State $G=G \setminus P$
		\EndWhile
		\EndFor
		\If{\texttt{$h(G)=0$}}

		\If{$t_B>0$}
		\State select any $v \in V(G)$
		\State $S=S\cup \{v\}$ and \texttt{Return S}
		\Else
		\State $S=S\cup \{v | v\in G \texttt{ AND } l(v)=1\}$ and \texttt{Return S}
		\EndIf
		\EndIf
	\end{algorithmic}
\end{algorithm}

Next we prove the correctness of Algorithm \ref{alg:minVEDB}. Note that, the modification of the labels of the reduced graph is done in such a way that if $l(v)=1$ for some $v\in V$, then $m(xy)=0$ for all edges incident to any vertex $x \in N_G(v)$. Hence we have the following claim:
\begin{cl}
After each iteration, in any block $B$, the set of edges with label 1 forms a clique.
\end{cl}
\begin{proof}
We will prove this by showing that it is not possible to have an edge $uv$ with $m(uv)=0$ and $m(ux)=m(vy)=1$ in any block. Suppose a block $B$ is having an edge $uv$ with $m(uv)=0$. Since $m(uv)=0$, there must be a vertex $p \in N[u]$ or $p \in N[v]$ with $l(p)=1$.

\begin{ca}
    (When $p \in N[u]$) : \textup{In this case all the edges incident to $u$ must be labeled $0$ since they are being ve-dominated by the vertex $p$. But $m(ux)=1$ for the edge $ux$, which is a contradiction.}
\end{ca}
\begin{ca}
    (When $p \in N[v]$) : \textup{In this case all the edges incident to $v$ must be labeled $0$ since they are being ve-dominated by the vertex $p$. But $m(vy)=1$ for the edge $vy$, which is a contradiction.}
\end{ca}
	Hence, if $m(uv)=0$ then either all edges incident to $u$ is labelled $0$ or all edges incident to $v$ is labelled $0$. And hence whenever an edge is labelled $0$ all other edges incident to at least one of its end point is also labelled $0$. It reduces the size of clique (formed by label-1 edges) by at least $1$.
\end{proof}
	
\begin{lem}
\label{lem:MinVEDBa}
Let $G$ be a block graph with an end block $B_0$ as root and $B$ be another end block such that $dist_G(B_0,B)=h(G)$. Also assume that $P=V(B)\setminus \{c\}$, where $c$ is the cut vertex of $B$ and $t_B$ denotes the number of edges with label 1 in $B$. Then followings are true.

(a) If $t_B\geq 2$, and $G'$ is new block graph results from $G$ by relabelling $c$ as $l(c)=1$, deleting all $v\in P$ and relabelling all edges $xy$ as $m(xy)=0  $ $\forall x\in N(c)$. Then $\gamma_{opve}(G)=\gamma_{opve}(G')$. 

(b) If $t_B=1$ but the edge with label-1 is not incident to $c$ and $G'$ is new block graph results from $G$ by relabelling $c$ as $l(c)=1$, deleting all $v\in P$ and relabelling all edges $xy$ as $m(xy)=0$ $\forall x\in N(c)$. Then $\gamma_{opve}(G)=\gamma_{opve}(G')$. 

(c) Let the conditions in (a) and (b) are not satisfied but $t_B=1$ and $G'$ is new block graph results from $G$ by relabelling $F(c)$ as $l(F(c))=1$, deleting all $v\in P$ and relabelling all edges $xy$ as $m(xy)=0  $ $\forall x\in N_{G'}(F(c))$. Then $\gamma_{opve}(G)=\gamma_{opve}(G')$.

(d) If $t_B=0$ and $B$ has $k$ many vertices with $l(v)=1$ and $G'$ is new block graph results from $G$ by deleting all $v\in P$. Then $\gamma_{opve}(G)=\gamma_{opve}(G') + k$. 

\end{lem}

\begin{proof}
(a)    Let $S$ be $\gamma_{opve}$-set  of $G$. If $\exists v \in P\cap S$ then $(S\setminus \{v\}) \cup \{c\}$ is also optional ve-dominating set of $G$, where $c$ is considered as a vertex with label $1$. So assume $S\cap P = \phi$. Now pick any edge $e$ with $m(e)=1$ in $G'$. There must be some $v\in S$ such that $v$ ve-dominates $e$. Also $v\not \in P$. Therefore $S$ is optional ve-dominating set of $G'$. Hence, $\gamma_{opve}(G') \le \gamma_{opve}(G)$.
			
Conversely, let $S'$ be $\gamma_{opve}$-set of $G'$. Since $l(c)=1, c\in S'$. Pick any edge $e$ with $m(e)=1$ from $G$. If $e\not \in B$ then obviously some $v\in S'$ ve-dominates $e$. If $e \in B$ then $c$ ve-dominates $e$. So, $S'$ is also optional ve-dominating set of $G$. Hence  $\gamma_{opve}(G) \le \gamma_{opve}(G')$.

(b)    The proof is same as the proof in (a).

(c)    Let $S$ be $\gamma_{opve}$-set of $G$. If $\exists v \in P\cap S$ then $(S\setminus \{v\}) \cup \{F(c)\}$ is also optional ve-dominating set of $G$. Since all edges private to $v$ are also ve-dominated by $F(c)$ 
    .  So assume $S\cap P = \phi$. Now pick any edge $e$ with $m(e)=1$ from $G'$. Since $S\cap P = \phi$ there must be some $u\in S$ such that $u$ ve-dominates $e$. Also $u\not \in P$. Therefore, $S$  is optional ve-dominating set of $G'$. Hence, $\gamma_{opve}(G') \le \gamma_{opve}(G)$.
			
    Conversely, let $S'$ be $\gamma_{opve}$-set of $G'$. Since $l(F(c))=1, F(c)\in S'$. So, all edges incident to $c$ is ve-dominated by $F(c)$. In block $B$ only one edge is labelled $1$ and is incident to $c$. So, it is ve-dominated by $F(c)$ and $S'$ is optional ve-dominating set of $G$. Hence,  $\gamma_{opve}(G) \le \gamma_{opve}(G')$.

(d)    Let $S$ be $\gamma_{opve}$-set  of $G$. $Q=\{ p|p\in P \text{ and } l(p)=1\}$. So $Q\subseteq S$. There are two cases $Q = \phi $ and $Q \not = \phi $.

\begin{ca}
	 ($Q = \phi $ $i.e.$ $(k=0)$) Pick any edge $e$ with $m'(e)=1$ from $G'$. Since $S$ is $\gamma_{opve}$-set of $G$. There must exist some $v\in S$ such that $v$ ve-dominates $e$ and this $v\not\in P$. Therefore, $\gamma_{opve}(G') \le \gamma_{opve}(G)$ and hence $\gamma_{opve}(G') \le \gamma_{opve}(G)$.
	
	Conversely, let $S'$ be $\gamma_{opve}$-set of $G'$. All the edges of $G$, except the edges of the newly added block $B$ is ve-dominated by $S'$ and none of the edges of block $B$ needs to be ve-dominated. Hence $\gamma_{opve}(G)\le \gamma_{opve}(G')$.
\end{ca}
\begin{ca}
    ($Q \not = \phi$ $i.e.$ $(k>0)$) Pick any edge $e$ with $m'(e)=1$ from $G'$. Since $Q\not=\phi$, $e$ is not incident to $c$. There must be some vertex $v \in S\setminus Q$ to ve-dominate $e$. Hence, $S\setminus Q$ is optional ve-dominating set of $G'$. Hence $\gamma_{opve}(G') \le \gamma_{opve}(G) - k$.
			
	Conversely, let $S'$ be $\gamma_{opve}$-set of $G'$. All the edges of $G$, except the edges of the newly added block $B$ is ve-dominated by $S'$ and none of the edges of block $B$ needs to be ve-dominated. Since $Q$ contains vertices with label $1$, $S'\cup Q$ is optional ve-dominating set of $G$. Hence, $\gamma_{opve}(G)\le\gamma_{opve}(G') + k$.	     
\end{ca}
\end{proof}

\begin{lem}\label{lem:MinVEDBb}
	Let $G$ be a complete graph, $i.e.$, $G=B$. If $t_B\geq 1$ , then $\gamma_{opve}(G)=1$. Otherwise, $\gamma_{opve}(G)=k$, where $k$ is the number of vertices of $B$ with $l(v)=1$.
\end{lem}
\begin{proof}
	When $t_B\geq 1$ then $B$ does not have any vetex with label-$1$. So, we need at least one vertex from block $B$ to ve-dominate all the edges with $m(e)=1$ and only one vertex is sufficient to ve-dominate all the edges. Hence,  $\gamma_{opve}(G)=1$.
	
	When $t_B = 0$, none of the edges needs to be ve-dominated. So, all the vertices with $l(v)=1$ forms an optional ve-dominating set and there are $k$ many such vertices. Hence, $\gamma_{opve}(G)=k$.
\end{proof}	
Lemma \ref{lem:MinVEDBa} and Lemma \ref{lem:MinVEDBb} shows that the output of Algorithm \ref{alg:minVEDB} is minimum optional ve-dominating set. At each iteration, we are taking $\mathcal{O}(deg(c))$ time. Hence the running time of Algorithm \ref{alg:minVEDB} is $\mathcal{O}((n + m)$. Thus, we have the following theorem.

\begin{theo}
The ve-domination problem can be solved in $\mathcal{O}(n + m)$ time for block graphs.
\end{theo}

\begin{remk}
We can also find the minimum independent ve-dominating set of a given block graph in a similar approach. The algorithm is similar to Algorithm \ref{alg:minVEDB}, but with little modification, to ensure the output is an independent set. Thus, for block graphs, the independent ve-domination problem can also be solved in linear time.
\end{remk}
\section{Undirected Path Graphs}
\label{sec:UDPgraph}

In this section, we prove that the ve-domination problem for undirected path graphs is NP-complete by showing a polynomial time reduction from $3$-dimensional matching problem which is a well-known NP-complete problem \cite{GareyJohnson1990}. 
A graph $G$ is called an \emph{undirected path graph} if $G$ is the intersection graphs of a family of paths of a tree.
In \cite{Gavril1975}, Gavril proved that a graph $G=(V,E)$ is an undirected path graph if and only if there exists a tree $T$ whose vertices are the maximal cliques of $G$ and the set of all maximal cliques containing a particular vertex $v$ of $V$ forms a path in $T$. This tree is called the clique tree of the undirected path graph $G$. The $3$-dimensional matching problem is as follows: given a set $M\subseteq U\times V\times W$, where $U,V$ and $W$ are disjoint set with $|U|=|V|=|W|=q$, does $M$ contains a matching $M'$, i.e., a subset $M'\subseteq M$ such that $|M'|=q$ and no two elements of $M'$ agree in any coordinate?

\begin{theo}
The ve-domination problem is NP-complete for undirected path graphs.
\end{theo}
\begin{proof}
It is easy to see that ve-domination problem is in NP. Now, we describe polynomial reduction form $3$-dimensional matching problem to ve-domination problem in undirected path graph. Let $U=\{u_r| 1\leq r\leq q\}, V=\{v_s| 1\leq s\leq q\}, W=\{w_t|1\leq t\leq q\}$, and $ M=\{m_i=(u_r,v_s,w_t)| 1 \leq i \leq p, u_r\in U, v_s\in V, w_t\in W\}$ be an instance of $3$-dimensional matching problem.
Now we construct a tree $T$ having $8p+6q+1$ vertices that becomes the clique tree of an undirected path graph $G$. The vertices of the tree $T$ are maximal cliques of $G$. The vertex set and the edge set are as follows:

For $1\leq i\leq p$, each $m_i=(u_r,v_s,w_t)\in M$ corresponds to $8$ cliques which are vertices of $T$, namely $\{A_i,B_i,C_i,D_i \}$, $\{A_i,B_i,D_i,F_i \}$, $\{C_i,D_i,G_i\}$, $\{A_i,B_i,E_i\}$, $\{C_i,G_i,K_i\}$, $\{A_i,E_i,H_i\}$, $\{B_i,E_i,I_i\}$, and $\{B_i,I_i,J_i\}$. These vertices depend only on the triple $m_i$ itself but not on the elements within the triple. These eight vertices induces a subtree corresponding to $m_i$ as illustrated in Figure \ref{fig:NPC_in_UPG}. Further, for each $u_r\in U$, $1\leq r\leq q$, we take two cliques $\{R_r\}\cup \{A_i|u_r\in m_i\}$ and $\{R_r, X_r\}$ which are vertices of $T$ forming a subtree as shown in Figure \ref{fig:NPC_in_UPG}. Similarly, for each $v_s\in V$, $1\leq s\leq q$ and $w_t\in W$, $1\leq t\leq q$, we add the cliques $\{S_s\} \cup \{B_i | v_s \in m_i \}$,$\{S_s, Y_k\}$ and $\{T_t\} \cup \{C_i | w_t \in m_i\}$, $\{T_t, Z_t\}$, respectively to the tree $T$ as shown in Figure \ref{fig:NPC_in_UPG}.Finally, $\{A_i,B_i,C_i | 1 \le i \le p\}$ is the last vertex of tree $T$. The construction of $T$ is illustrated in Figure \ref{fig:NPC_in_UPG}.

\begin{figure}[h]
	\begin{center}
	\includegraphics[scale=0.6]{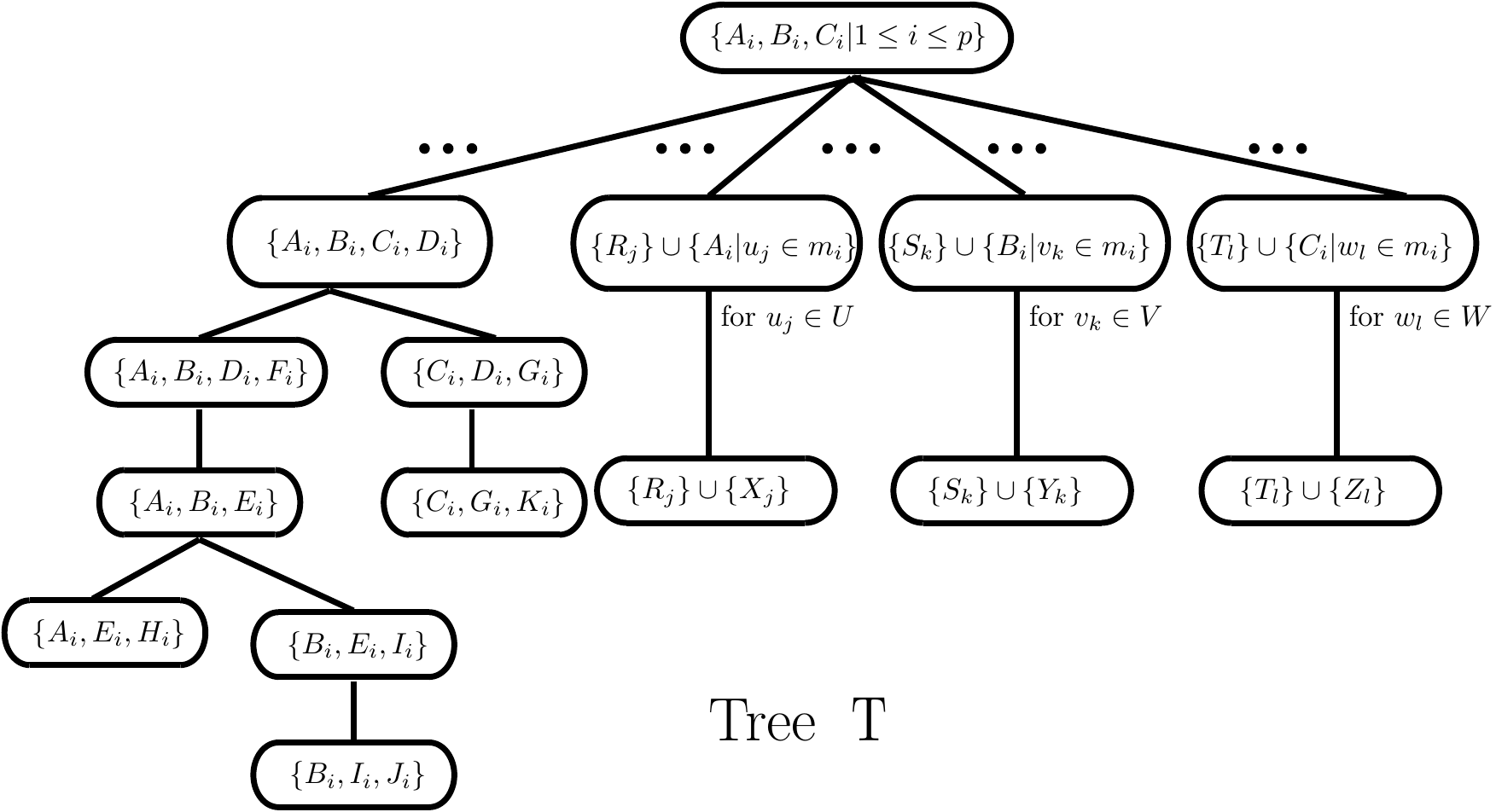}
	\end{center}
	\caption{The clique tree of the undirected path graphs}
    \label{fig:NPC_in_UPG}
\end{figure}

Hence, $T$ is the clique tree of the undirected path graph $\mathcal{G}$ whose vertex set is $$\{A_i,B_i,C_i,D_i,E_i,F_i,G_i,H_i,I_i,J_i,K_i | 1 \le i \le p\} \cup \{R_j,S_j,T_j,X_j,Y_j,Z_j: 1\le j\le q\}.$$

\begin{cl}
The graph $\mathcal{G}$ has a ve-dominating set of size $2p + q$ if and only if $3$-dimensional matching has a solution.
\end{cl}
\begin{proof}
Let $\mathcal{D}$ be a ve-dominating set of $\mathcal{G}$ of size $2p+q$. For any $i\in \{1,2,\ldots, p\}$, the only way to ve-dominate the edge-set of the subgraph induced by the vertex set $\{A_i,B_i,C_i,D_i,E_i,F_i,G_i,H_i,I_i,J_i,K_i\}$ corresponding to $m_i$ with two vertices is to choose $D_i$ and $E_i$. Hence, to ve-dominate the edge-set of that induced subgraph by any larger vertex set, at least three vertices has to be taken. Note that, the set $\{A_i,B_i, C_i\}$ ve-dominates the edge-set of that induced subgraph. So, without loss of generality, we assume that $\mathcal{D}$ consists of $A_i,B_i,C_i$ for $t$ many $m_i$'s and $D_i,E_i$ for $(p-t)$ many other $m_i$'s. Also to ve-dominate the edges of the form $R_rX_r$, $S_sY_s$ and $T_tZ_t$, $\mathcal{D}$ contains at least $max\{3(q-t),0\}$ many vertices (namely, $R_j$ or $X_j$,$S_k$ or $Y_k$,$T_l$ or $Z_l$). Hence, we have,

$$2p+q=|\mathcal{D}| \geq 3t + 2(p-t) + 3(q-t)= 2p +3q -2t.$$

So, $t \geq q$. i.e. $\mathcal{D}$ must contain at least $q$ many $A_i,B_i,C_i$. Picking the corresponding $m_i$'s form a matching $M'$ of size $q$.

Conversely, let $M'$ be the solution of the $3$-dimensional matching problem of size $q$. Then we can form the ve-dominating set $\mathcal{D}$ as $\mathcal{D} = \{A_i,B_i,C_i : m_i \in M'\} \cup \{D_i,E_i : m_i \not\in M'\}$. Clearly, $\mathcal{D}$ is a ve-dominating set of $\mathcal{G}$ of size $2p+q$. 
\end{proof}
Hence, the ve-domination problem is NP-complete for undirected path graph. 
\end{proof}
   
\section{Trees with Equal $\gamma_{ve}$ and $i_{ve}$}
\label{sec:VEequalIVE}

For every graph $G$, the independent ve-domination number is obviously at least as large as the ve-domination number. In this section, we characterize the trees for which these two parameters $\gamma_{ve}$ and $i_{ve}$ are equal. We start with some pertinent definitions.

\begin{defi}
An \emph{atom} $A$ is a tree with at least $3$ vertices with a vertex, say $c$, designated as center of the atom such that distance of every vertex from $c$ is at most $2$.
\end{defi}

We denote an atom along with its center by $(A,c)$. Note that the center $c$ ve-dominates all edges of the atom $A$. Next, we define an operation for joining two atoms to construct a bigger tree.
\begin{defi}
Let $(A',c')$ and $(A,c)$ be two atoms along with their centers $c'$ and $c$, respectively. For some $i,j\in \{0,1,2\}$, we define \emph{$(i-j)$-join}, denoted by $\bigotimes$, as the addition of an edge $(x_{c'},x_{c})$ between the vertices $x_{c'}\in V(A')$ and $x_{c}\in V(A)$ such that $dist_{A'}(c',x_{c'})=i$ and $dist_{A}(c,x_c)=j$. 
\end{defi}

With slight abuse of notation, $T'\bigotimes (A,c)$ denotes the tree obtained by $(i-j)$-join between two atoms $(A',c')$ and $(A,c)$, where $(A',c')$ is an atom in $T'$. Given a subset $S \subseteq V$, a vertex $v\in S$ has a \emph{private edge} $e \in E$ with respect to the set $S$ if $v$ ve-dominates the edge $e$ and no other vertex in $S$ ve-dominates the edge $e$. An edge $e=xy \in E$ is called a \emph{distance-$1$ private edge} of $v\in S$ with respect to the set $S$ if $e$ is a private edge of $v$ with respect to $S$ and $\min\{dist(v,x),dist(v,y)\}=1$. Next, we give the recursive definition of a family of trees, say $\mathscr{T}$, using the notion of atom and $(i-j)$-join. 

\begin{defi}\label{defi:scriptT}
The recursive definition of the family $\mathscr{T}$ of trees is as follows:

\begin{enumerate}
\item every atom $(A,c)\in \mathscr{T}$ and

\item Let $T'\in \mathscr{T}$ and $(A',c')$ be an atom in $T'$ and $S'$ be the set of all atom centers in $T'$. Then $T=T'\bigotimes (A,c) \in \mathscr{T}$ if one of the following cases hold

        \begin{enumerate}
            \item[$(i)$] $\bigotimes$ is a $(0-1)$-join such that
            \begin{enumerate}
                \item[$(a)$] $c'$ has a neighbour $y$ such that all edges incident to $y$, except $yc'$, are pendent edges. Also, $c$ has no distance-$1$ edge and at least two edges of $(A,c)$ is incident to $c$. 
               
                \item[$(b)$] $c'$ has distance-$1$ private edges with respect to $S'$ and at least one distance-$1$ edge of $c$ is not incident to $x_c$. 
                
                \item[$(c)$] $c'$ has no distance-$1$ private edge  with respect to $S'$ and at least one distance-$1$ edge of $c$ is not incident to $x_c$. 
               
            \end{enumerate}
            \item[$(ii)$] $\bigotimes$ is a $(1-0)$-join such that
           \begin{enumerate}
                 \item[(a)] $c$ has distance-$1$ edges and $c'$ has a neighbour $y$ such that all edges incident to $y$, except $c'y$, are pendent and private edges of $c'$ with respect to $S'$. 
                
                 \item[(b)] $c$ has distance-$1$ edges and $c'$ has a neighbour $y$ such that $y$ is a leaf vertex  and $y\not = x_{c'}$. 
                
            \end{enumerate}
            \item[$(iii)$] $\bigotimes$ is a $(1-1)$-join when at least one distance-$1$ private edge of $c'$ with respect to $S'$ is not incident to $x_{c'}$ and at least one distance-$1$ edge of $c$ is not incident to $x_c$. 
            
            \item[$(iv)$] $\bigotimes$ is a $(2-1)$-join when all distance-$1$ private edges of $c'$ with respect to $S'$ are not incident to $y$, where $y\in N_{T'}(c') \cap N_{T'}(x_{c'})$, and at least one distance-$1$ edge of $c$ is not incident to $x_c$. 
            
        \end{enumerate}
    \end{enumerate}
\end{defi}

Now, we show that if $T\in \mathscr{T}$, then \emph{ve-domination number} and \emph{independent ve-domination number} are same.

\begin{lem}
	\label{lem:atomCenterVED}
	If $T\in \mathscr{T}$, then the set of all atom centers of $T$ forms a minimum ve-dominating set.
\end{lem}

\begin{proof}
	We prove this by induction on the number of atoms in $T$. Clearly, when $T$ is an atom, the hypothesis is true. Let $T \in \mathscr{T}$ be a tree containing $k$ atoms and $T$ is obtained from $T'\in \mathscr{T}$ by joining the atom $(A,c)$ with an atom $(A',c')$ of $T'$ satisfying the joining rules. Let $S$ and $S'$ be the atom centers of $T$ and $T'$, respectively. Clearly, $S=S' \cup \{c\}$. By induction hypothesis, $S'$ is a $\gamma_{ve}$-set of $T'$. We show that $S$ is a $\gamma_{ve}$-set of $T$ for all the seven types of joining defined in Definition \ref{defi:scriptT}.
	
\begin{description}
\item[\textbf{${\bf(0-1)(a):}$}] In this type of joining, $(A,c)$ is star and $c'$ has a neighbour $y$ such that all edges incident to $y$, except $yc'$, are pendent edges. Since $S'$ is a ve-dominating set of $T'$, $S$ is obviously a ve-dominating set of $T$. If possible, let us assume that $S$ is not a $\gamma_{ve}$-set of $T$. Let $D$ be a $\gamma_{ve}$-set of $T$ such that $|D|< |S|$. Note that $D$ contains exactly one vertex from $(A,c)$, say $p$, to ve-dominate all edges of $(A,c)$. Also, to ve-dominate the pendent edges which are incident to $y$, $D$ must contain one vertex, say $q$, from $\{x,y,c'\}$, where $x$ is any leaf adjacent to $y$. Since the set of edges that are ve-dominated by $\{p,q\}$ can also be ve-dominated by $\{c,c'\}$, $D'= (D\setminus \{p,q\})\cup \{c,c'\}$ is also a $\gamma_{ve}$-set of $T$. It is easy to see that $D' \setminus \{c\}$ is a $\gamma_{ve}$-set of $T'$. This contradicts the fact that $S'$ is a ve-dominating set of $T'$ of minimum cardinality. Hence, $S$ is a $\gamma_{ve}$-set of $T$.
		
		\item[\textbf{${\bf(0-1)(b):}$}] If possible, let us assume that $S$ is not a $\gamma_{ve}$-set of $T$. Let $D$ be a $\gamma_{ve}$-set of $T$ such that $|D|< |S|$. Clearly, $D$ contains exactly one vertex, say $p$, from $(A,c)$. Since the set of edges that are ve-dominated by $p$ can also be ve-dominated by $c$, $D'=(D\setminus \{p\})\cup \{c\}$ is also a $\gamma_{ve}$-set of $T$. It is easy to see that $D' \setminus \{c\}$ is a $\gamma_{ve}$-set of $T'$. This is a contradiction. Hence, $S$ is a $\gamma_{ve}$-set of $T$.
		
		\item[\textbf{${\bf(0-1)(c):}$}] If possible, let us assume that $S$ is not a $\gamma_{ve}$-set of $T$. Let $D$ be a $\gamma_{ve}$-set of $T$ such that $|D|< |S|$. Clearly, $D$ contains exactly one vertex, say $p$, from $(A,c)$. Since the set of edges that are ve-dominated by $p$ can also be ve-dominated by $c$, $D'=(D\setminus \{p\})\cup \{c\}$ is also a $\gamma_{ve}$-set of $T$. It is easy to see that $D' \setminus \{c\}$ is a $\gamma_{ve}$-set of $T'$. This is a contradiction. Hence, $S$ is a $\gamma_{ve}$-set of $T$.
		
		\item[\textbf{${\bf(1-0)(a):}$}] In this type of joining, $c'$ has a neighbour $y$ such that all edges incident to $y$, except $(yc')$, are pendent edges. If possible, let us assume that $S$ is not a $\gamma_{ve}$-set of $T$. Let $D$ be a $\gamma_{ve}$-set of $T$ such that $|D|< |S|$. Note that $D$ contains exactly one vertex from $(A,c)$, say $p$, to ve-dominate all edges of $(A,c)$. Suppose $x$ is a leaf node adjacent to $y$. So, to ve-dominate the pendent edge $xy$, $D$ must contain one vertex, say $q$, from $\{x,y,c'\}$. Since the set of edges that are ve-dominated by $\{p,q\}$ can also be ve-dominated by $\{c,c'\}$, $D'= (D\setminus \{p,q\})\cup \{c,c'\}$ is also a $\gamma_{ve}$-set of $T$. It is easy to see that $D' \setminus \{c\}$ is a $\gamma_{ve}$-set of $T'$. This is a contradiction. Hence, $S$ is a $\gamma_{ve}$-set of $T$.
		
		\item[\textbf{${\bf(1-0)(b):}$}] In this case, $c'$ has a neighbour $y \not = x_{c'}$ such that $y$ is leaf vertex. If possible, let us assume that $S$ is not a $\gamma_{ve}$-set of $T$. Let $D$ be a $\gamma_{ve}$-set of $T$ such that $|D|< |S|$. Note that $D$ contains exactly one vertex from $(A,c)$, say $p$, to ve-dominate all edges of $(A,c)$. Since $p$ cannot ve-dominate the edge $c'y$, $D$ must contain a vertex, say $q$, to ve-dominate this edge. Note that $q\in N[c']$ because $y$ is leaf vertex and $q$ also ve-dominate the edge $c'x_{c'}$. Since the set of edges that are ve-dominated by $p$ can also be ve-dominated by $c$, $D'= (D\setminus \{p\})\cup \{c\}$ is also a $\gamma_{ve}$-set of $T$. It is easy to see that $D' \setminus \{c\}$ is a $\gamma_{ve}$-set of $T'$
		. This is a contradiction. Hence, $S$ is a $\gamma_{ve}$-set of $T$.
		
		\item[\textbf{${\bf(1-1)}$-join:}] Since $S'$ is a ve-dominating set of $T'$, $S$ is obviously a ve-dominating set of $T$. If possible, let us assume that $S$ is not a $\gamma_{ve}$-set of $T$. Let $D$ be a $\gamma_{ve}$-set of $T$ such that $|D|< |S|$. Clearly, $D$ contains exactly one vertex, say $p$, from $(A,c)$. Since the set of edges that are ve-dominated by $p$ can also be ve-dominated by $c$, $D'=(D\setminus \{p\})\cup \{c\}$ is also a $\gamma_{ve}$-set of $T$. It is easy to see that $D' \setminus \{c\}$ is a $\gamma_{ve}$-set of $T'$. This is a contradiction. Hence, $S$ is a $\gamma_{ve}$-set of $T$.
		
		\item[\textbf{${\bf(2-1)}$-join:}]  Since $S'$ is a ve-dominating set of $T'$, $S$ is obviously a ve-dominating set of $T$. If possible, let us assume that $S$ is not a $\gamma_{ve}$-set of $T$. Let $D$ be a $\gamma_{ve}$-set of $T$ such that $|D|< |S|$. Clearly, $D$ contains exactly one vertex, say $p$, from $(A,c)$. Since the set of edges that are ve-dominated by $p$ can also be ve-dominated by $c$, $D'=(D\setminus \{p\})\cup \{c\}$ is also a $\gamma_{ve}$-set of $T$. It is easy to see that $D' \setminus \{c\}$ is a $\gamma_{ve}$-set of $T'$. This is a contradiction. Hence, $S$ is a $\gamma_{ve}$-set of $T$. 
\end{description}
\end{proof}

\begin{theo}\label{theo:gammave=ive}
	For all $T \in \mathscr{T}$, $\gamma_{ve}(T)=i_{ve}(T)$.
\end{theo}
\begin{proof}
	By the definition of $\mathscr{T}$, the distance between any two atom centers in $T$ is at least $2$. Hence, the set of these atom centers, say $S$, forms an independent set. In Lemma \ref{lem:atomCenterVED}, $S$ is also a $\gamma_{ve}$-set. Hence, $\gamma_{ve}(T)=i_{ve}(T)$ for all $T \in \mathscr{T}$. 
\end{proof}

Next, we show that the converse of Theorem \ref{theo:gammave=ive} is also true. For that, first we prove following lemmas that allow us to construct an \emph{independent ve-dominating set} of a tree with some desirable properties.

\begin{lem}\label{lem:noLeafIVE}
For any tree $T$ $(n\geq 3)$, there exists an $i_{ve}$-set which does not contain any leaf.
\end{lem}
\begin{proof}
Let $S$ be an $i_{ve}$-set of $T$. If $S$ does not contain any leaf, then we are done. Otherwise assume that $S$ contains a leaf, say $x$. Let the neighbour of $x$ is $y$ and $N(y)=\{z_1,z_2,\ldots,z_q,x\}$. Since $S$ is independent set, $y\not \in S$. Also, none of the $z_i$ are in $S$, because if any of $z_i \in S$, then $S\setminus \{x\}$ is also an independent ve-dominating set. Hence by replacing $x$ by $y$, we get another $i_{ve}$-set. Repeating this process, we can form an $i_{ve}$-set of $T$ which does not contain any leaf.
\end{proof}

\begin{lem}\label{lem:dist3-4IVE}
Let $S$ be an $i_{ve}$-set of a rooted tree $T$ having depth $h$, which does not have any leaf. If the vertex $u\in S$ is at $(h-1)^{th}$-level and $v$ is the closest vertex to $u$ such that $v\in S$ and $dist(u,v)\geq 3$, then $S'=(S\setminus \{u\})\cup \{w\}$ is also an $i_{ve}$-set without having any leaf, where $w$ is the parent of $u$ in $T$.
\end{lem}
\begin{proof}
It is easy to see that all the edges that are ve-dominated by $u$ can also be ve-dominated by $w$. So, $S'$ is $\gamma_{ve}$-set. Also, since the minimum distance between $u$ and $v$ is at least $3$, $S'$ is independent set. Hence, $S'$ is an $i_{ve}$-set without having any leaf.  
\end{proof}


\begin{theo}\label{theo:TinmathscrT}
	If $\gamma_{ve}(T)=i_{ve}(T)$ for a tree $T$ with $n\geq 3$, then $T\in \mathscr{T}$.
\end{theo}
\begin{proof}
	We prove this by induction on the size of $i_{ve}$-set. As a base case when $i_{ve}(T)=1$ then $T$ is an atom, hence $T\in \mathscr{T}$. Assume the induction hypothesis, if $\gamma_{ve}(T)=i_{ve}(T)=k-1$ then $T\in \mathscr{T}$ and all the elements of $i_{ve}$-set represent centers of the underlying atoms of $T$. 
	
	Let a tree $T_v$ is rooted at a vertex $v$ and have $h$ levels as $1, 2, 3, \ldots,h$ such that root is at $1^{st}$-level and most distant leaf from root is at $h^{th}$-level. Let us also assume that $S$ is an $i_{ve}$-set of the tree $T_v$  such that $|S|=k$. Clearly, $S$ has no vertices from $h^{th}$-level(Lemma \ref{lem:noLeafIVE}). Consider a vertex $c \in S$ that is at maximum distance from the root (if more than one vertices are at the same maximum level then consider any one who has descendent at $h^{th}$-level). So, $c$ is either at $(h-1)^{th}$-level or at $(h-2)^{th}$-level. Let $c'$ is the nearest vertex to $c$ such that $c' \in S$. So, $dist(c,c')=2,3$ or $4$, otherwise, $S$ will not be an $i_{ve}$-set. Now, we extract an atom $A$ with center $c$ to form a tree $T'$. Assume $S'$ be the $i_{ve}$-set and $\gamma_{ve}$-set of $T'$. By induction hypothesis $T'\in \mathscr{T}$ and $S'$ is set of all atom centers of $T'$. Clearly, $c'$ is center of an atom, say $A'$, in $T'$. We show that the joining of atom $A$ with tree $T'$ follows the joining rules defined in Definition \ref{defi:scriptT}. The whole proof is divided into three cases depending on the distance between $c$ and $c'$. Further, each case is divided into two subcases considering $c$ can be at $(h-1)^{th}$-level or at $(h-2)^{th}$-level. These different exhaustive cases are explored as follows:
	
	\begin{ca}\label{cas:dist2} : $dist(c,c')=2$ 
		\begin{enumerate}
			\item[(A)] $c$ is at $(h-1)^{th}$-level : 
			
			\emph{Since $dist(c,c')=2$, $c'$ can either be at $(h-1)^{th}$-level or at $(h-3)^{th}$-level. Both of these possibilities are considered as follows:}\\
			\begin{enumerate}
				\item[(i)] \texttt{$c'$ is at $(h-1)^{th}$-level} : \emph{This case is not possible because we can find a new set $D=S\setminus\{c,c'\}\cup\{parent(c)\}$ such that $D$ is $i_{ve}$-set and $|D| < |S|$ which violates the minimality of $S$.}
				\item[(ii)] \texttt{$c'$ is at $(h-3)^{th}$-level} : \emph{Only two possibilities are there, $c'$ has some distance-$1$ private edges with respect to $S$ or it has no distance-$1$ private edge with respect to $S$. Both of these cases are considered in following three points:}\\
				\begin{enumerate}
					\item[(a)] $c'$ has a neighbour $y$ such that all other edges incident to $y$ are pendent edges : \emph{In this case, all edges incident to $y$, except $c'y$, are distance-$1$ private edges of $c'$ with respect to $S$. Suppose $A$ is a sub-tree rooted at $parent(c)$, we can think $A$ as an atom with center $c$. Remove $A$ from $T$ to form tree $T'$. Assume $c'$ is the center of an atom $A'$ in $T'$. Now, it is very easy to see that the joining between $(A,c)$ and $(A',c')$ resembles the $0-1$ joining of type $(a)$  defined in Definition \ref{defi:scriptT}}.
					
					\item[(b)] $c'$ has no neighbour $y$ such that all other edges incident to $y$ are pendent edges, but it has distance-$1$ private edges with respect to $S$ : \emph{In this case, edges incident to $x'=parent(c')$ are distance-$1$ private edges of $c'$ with respect to $S$. Clearly, none of the neighbours of $x'$, except $c'$, is in $S$. So, $D=S\setminus \{c,c'\} \cup\{x=parent(c),x'\}$ is $\gamma_{ve}$-set as well as $i_{ve}$-set with $x'$ having no distance-$1$ private edges. Suppose $A$ is the sub-tree rooted at vertex $c'$, we can think $A$ as an atom with center $x$. Remove $A$ from $T$ to form $T'$. Assume $x'$ is the center of an atom $A'$ in $T'$. Now, it is easy to see that the joining between $(A,x)$ and $(A',x')$ resembles the $0-1$ joining of type $(c)$  defined in Definition \ref{defi:scriptT}}.
					\item[(c)] $c'$ has no distance-$1$ private edges with respect to $S$ : \emph{This case does not arise because all the private edges of $c'$ and $c$ are ve-dominated by $parent(c)$. So, $S$ is not the minimal independent ve-dominating set}.
				\end{enumerate}
			\end{enumerate}
			
			\item[(B)] $c$ is at $(h-2)^{th}$-level :
			
			\emph{Since $dist(c,c')=2$, $c'$ can either be at $(h-2)^{th}$-level or at $(h-4)^{th}$-level. Both of these cases are considered as follows:}\\
			\begin{enumerate}
				\item[(i)] \texttt{$c'$ is at $(h-2)^{th}$-level} : \emph{There are two possibilities, $c'$ has some distance-$1$ private edges with respect to $S$ or it has no distance-$1$ private edge with respect to $S$}.\\
				\begin{enumerate}
					\item[(a)] $c'$ has distance-$1$ private edge with respect to $S$ : \emph{Since $c$ and $c'$ both are at $(h-2)^{th}$-level, the distance-$1$ private edges of $c'$ must be pendent edges (incident to leaf vertices). $i.e.$ $c'$ has a neighbour $y$ such that all the edges incident to $y$, except $c'y$, are pendent edges. Suppose $A$ is the sub-tree rooted at vertex $c$, we can think $A$ as an atom with center $c$. Remove $A$ from $T$ to form tree $T'$. Assume $c'$ is the center of an atom $A'$ in $T'$. Now, it is very easy to see that the joining between $(A',c')$ and $(A,c)$ resembles the $1-0$ joining of type $(a)$ defined in Definition \ref{defi:scriptT}}.
					
					\item[(b)] $c'$ has no distance-$1$ private edge with respect to $S$: \emph{Since, $c'$ is at $(h-2)^{th}$-level and $dist(c,c')=2$, all the private edges of $c'$ with respect to $S$ must be pendent edges. Suppose $A$ is the sub-tree rooted at vertex $c$,  we can think $A$ as an atom with center $c$. Remove $A$ from $T$ to form tree $T'$. Assume $c'$ is the center of an atom $A'$ in $T'$. Now, the joining between $(A',c')$ and $(A,c)$ resembles the $1-0$ joining of type $(b)$ defined in Definition \ref{defi:scriptT}}.
				\end{enumerate}
				
				\item[(ii)] \texttt{$c'$ is at $(h-4)^{th}$-level} : \emph{Only two possibilities are there, either $c'$ has some distance-$1$ private edges or it has no distance-$1$ private edge with respect to $S$}.\\
				\begin{enumerate}
					\item[(a)] $c'$ has no distance-$1$ private edge with respect to $S$ : \emph{Since, $c' \in S$, $c'$ must have a private edge with respect to $S$. We have chosen $c$ in such a way that it has descendent at $h^{th}$-level, and edges incident to such descendents are distance-$1$ private edges of $c$  with respect to $S$. Suppose $A$ is the sub-tree rooted at vertex $x=parent(c)$,  we can think $A$ as an atom with center $c$. Remove $A$ from $T$ to form tree $T'$. Assume $c'$ is the center of an atom $A'$ in $T'$. Now, the joining between $(A,c)$ and $(A',c')$ resembles the $0-1$ joining of type $(c)$ defined in Definition \ref{defi:scriptT}}.
					\
					\item[(b)] $c'$ has distance-$1$ private edge with respect to $S$ : \emph{$c$ already has private edge at distance-$1$ with respect to $S$ and $dist(c,c')=2$. Suppose $A$ is the sub-tree rooted at vertex $x=parent(c)$, we can think $A$ as an atom with center $c$. Remove $A$ from $T$ to form $T'$. Assume $c'$ is the center of an atom $A'$ in $T'$. Now,  the joining between $(A,c)$ and $(A',c')$ resembles the $0-1$ joining of type $(b)$ defined in Definition \ref{defi:scriptT}}.
				\end{enumerate}
			\end{enumerate}
		\end{enumerate}
	\end{ca}
	\begin{ca}\label{cas:dist3} : $dist(c,c')=3$
		\begin{enumerate}
			\item[(A)] $c$ is at $(h-1)^{th}$-level : 
			
			\emph{Since $dist(c,c')=3$, $c'$ can either be at $(h-2)^{th}$-level or at $(h-4)^{th}$-level. In both of these cases we apply Lemma \ref{lem:dist3-4IVE} and distance between $c$ and $c'$ become $2$ which we already have considered in Case \ref{cas:dist2}}
			\item[(B)] $c$ is at $(h-2)^{th}$-level : 
			
			\emph{Since $dist(c,c')=3$, $c'$ can either be at $(h-3)^{th}$-level or at $(h-5)^{th}$-level. Both of these cases are considered as follows:}
			\begin{enumerate}
				\item[(i)] \texttt{$c'$ is at $(h-3)^{th}$-level} : \emph{Consider a vertex $x=parent(c')$. Only four possibilities are there, \textbf{(a)} $c'$ has some distance-$1$ private edges with respect to $S$ some of which are incident to $x$ and some are incident to any child $y$ of $c'$, \textbf{(b)} $c'$ has distance-$1$ private edges with respect to $S$ and none of these edges are incident to $x$, \textbf{(c)} $c'$ has distance-$1$ private edges with respect to $S$ all of which are incident to $x$ and \textbf{(d)} $c'$ has no distance-$1$ private edge with respect to $S$. All these four possibilities are explored as follows:}\\
				\begin{enumerate}
					\item[(a)]  $c'$ has some distance-$1$ private edges with respect to $S$ some which are incident to $x$ and some(at least one) are not incident to $x$: \emph{We have chosen $c$ in such a way that it has descendent at $h^{th}$-level, and edges incident to such descendents are private edges of $c$ at distance-$1$ with respect to $S$. Suppose $A$ is the sub-tree rooted at vertex $w=parent(c)$, we can think $A$ as an atom with center $c$. Remove $A$ from $T$ to form $T'$. Assume $c'$ is the center of an atom $A'$ in $T'$. Now, it is easy to see that the joining between $(A',c')$ and $(A,c)$ resembles the $1-1$ type joining defined in Definition \ref{defi:scriptT}}.
					\item[(b)] $c'$ has distance-$1$ private edges with respect to $S$ and none of these edges are incident to $x$ : \emph{As in previous case, here also we take $A$ as a sub-tree rooted at vertex $w=parent(c)$ and we can see that the joining between $(A',c')$ and $(A,c)$ resembles the $1-1$ joining defined in Definition \ref{defi:scriptT}}.
					
					\item[(c)] $c'$ has distance-$1$ private edges with respect to $S$ all of which are incident to $x$: \emph{Since an edge incident to $x$ is private edge of $c'$ with respect to $S$, none of the neighbours of $x$, other than $c'$ is in $S$. Also, all the private edges of $c'$ is also ve-dominated by $x$. Hence $D=S\setminus \{c'\}\cup \{x\}$ is also $\gamma_{ve}$-set as well as $i_{ve}$-set of $T$. Now, $dist(c,x)=2$ and this is already considered in Case \ref{cas:dist2}}.
					
					\item[(d)] $c'$ has no distance-$1$ private edge with respect to $S$: \emph{Since, $c' \in S$, $c'$ must have a private edge with respect to $S$. Clearly, all the private edges of $c'$ is incident to it. If none of the neighbours of $x$ except $c'$ is in $S$ then $D=(S\setminus\{c'\} \cup \{x\}$), is also $\gamma_{ve}$-set as well as $i_{ve}$-set of $T$. Now, $dist(c,x)=2$, where $x\in S'$. This possibility is already considered in Case \ref{cas:dist2}(B). If any other neighbour of $x$ is in $S$ then that vertex, say $w$, must have distance-$1$ private edge with respect to $S$. Here, $dist(c,w)=3$, and joining between $(A,c)$ and $(A',w)$ resembles $1-1$ joining defined in Definition \ref{defi:scriptT}}.
				\end{enumerate}  
				\item[(ii)] \texttt{$c'$ is at $(h-5)^{th}$-level} : \emph{Consider a vertex $x$ such that $x$ is child of $c'$ and $dist(c,x)=2$. Similar to the previous sub-case (sub-case (i)), we have four exhaustive possibilities.
					All these four possibilities are explored as follows:}\\
				\begin{enumerate}	    
					\item[(a)] $c'$ has distance-$1$ private edge(s) with respect to $S$ some of which are incident to $x$ and at least one of these edges are not incident to $x$ :  \emph{Its proof is almost same as the proof of $(a)$ in previous sub-case.
					}.
					
					\item[(b)] $c'$ has distance-$1$ private edges with respect to $S$ none of which are incident to $x$: \emph{Its proof is same as the proof of $(b)$ in previous sub-case.
					}.
					\item[(c)] $c'$ has distance-$1$ private edges with respect to $S$ all of which are incident to $x$ : \emph{Its proof is almost same as the proof of $(c)$ in previous sub-case.
					}.
					\item[(d)] $c'$ has no distance-$1$ private edge with respect to $S$: \emph{Proof is same as the proof of $(d)$ in previous sub-case.
					}.
				\end{enumerate}
			\end{enumerate}
		\end{enumerate}
	\end{ca}
	\begin{ca}\label{cas:dist4} : $dist(c,c')=4$
		\begin{enumerate}
			\item[(A)] $c$ is at $(h-1)^{th}$-level : 
			
			\emph{Since $dist(c,c')=4$ and $c$ is at $(h-1)^{th}$-level, after applying Lemma \ref{lem:dist3-4IVE} we can reduce the distance between $c$ and $c'$ to $3$ which we already have considered in Case \ref{cas:dist3}}.
			\item[(B)] $c$ is at $(h-2)^{th}$-level : 
			
			\emph{Since $dist(c,c')=4$, $c'$ can either be at $(h-2)^{th}$-level, $(h-4)^{th}$-level or at $(h-6)^{th}$-level. All of these cases are considered as follows:}
			\begin{enumerate}
				\item[(i)] \texttt{$c'$ is at $(h-2)^{th}$-level }: \emph{Consider a vertex $x=parent(c')$. Private edges of $c'$ can be one of the three types as follows:}\\
				\begin{enumerate}
					\item[(a)] At least one distance-$1$ private edge of $c'$ with respect to $S$ is not incident to $x$: \emph{Suppose $A$ is the sub-tree rooted at vertex $w=parent(c)$, we can think $A$ as an atom with center $c$. Remove $A$ from $T$ to form tree $T'$. Assume $c'$ is the center of an atom $A'$ in $T'$. Now, it is easy to see that the joining between $(A',c')$ and $(A,c)$ resembles the $2-1$ joining defined in Definition \ref{defi:scriptT}}.
					
					\item[(b)] All the distance-$1$ private edges with respect to $S$ are incident to $x$ : \emph{Since distance-$1$ private edges are incident to $x$, none of the neighbours of $x$, except $c'$ is in $S$. Hence, $D=S\setminus \{c'\}\cup \{x\}$ is also $\gamma_{ve}$-set as well as $i_{ve}$-set of $T$. Now, $dist(c,x)=3$ and this case is considered in Case \ref{cas:dist3}}.
					
					\item[(c)] $c'$ has no distance-$1$ private edge with respect to $S$: \emph{In this case, other than $c'$ at least one neighbour of $x$, say $y$, must be in $S$. Also notice that since $c'$ has no distance-$1$ private edge with respect to $S$, $y$ must have distance-$1$ private edge with respect to $S$.  Now, instead of $c$ and $c'$ consider two vertices $c$ and $y$. Suppose $A$ is the sub-tree rooted at vertex $w=parent(c)$, we can think $A$ as an atom with center $c$. Remove $A$ from $T$ to form tree $T'$. Assume $y$ is the center of an atom $A'$ in $T'$. Now, it is easy to see that the joining between $(A',y)$ and $(A,c)$ resembles the $2-1$ joining defined in Definition \ref{defi:scriptT}}.
				\end{enumerate}
				
				\item[(ii)] \texttt{$c'$ is at $(h-4)^{th}$-level }: \emph{Consider a vertex $x=parent(c')$. Private edges of $c'$ can be one of the three types as follows:}\\
				\begin{enumerate}
					\item[(a)] At least one distance-$1$ private edge of $c'$ with respect to $S$ is not incident to $x$: \emph{Proof is similar to the one we have seen in previous sub-case ($(a)$)
					}.
					\item[(b)] All the distance-$1$ private edges of $c'$ with respect to $S$ are incident to $x$ : \emph{First possibility is that all the private edges of $c'$ with respect to $S$ are incident to $x$. If this is the case then since nearest vertex to $c$ which is in $S$ is at distance $4$, none of the neighbours of $p$, where $p\in (\{u | u \text{ is }child(x)\} \cap \{u|dist(c,u)=2\})$, is in $S$ and hence $D=S\setminus \{c'\}\cup \{p\}$ is also $\gamma_{ve}$-set as well as $i_{ve}$-set of $T$. Now, $dist(c,p)=2$ and this case is considered in Case \ref{cas:dist2}. Other possibility is that some private edges of $c'$ with respect to $S$ is not incident to $x$. Since distance-$1$ private edges are incident to $x$, none of the neighbours of $x$, except $c'$ is in $S$. Hence, $D=S\setminus \{c'\}\cap \{x\}$ is also $\gamma_{ve}$-set as well as $i_{ve}$-set of $T$. Now, $dist(c,x)=3$ and this case is considered in Case \ref{cas:dist3}}.
					\item[(c)] $c'$ has no distance-$1$ private edge with respect to $S$: \emph{Explanation for this one is same as we have discussed in previous sub-case ($(c)$)}.
				\end{enumerate}
				\item[(iii)] \texttt{$c'$ is at $(h-6)^{th}$-level} : \emph{Consider a vertex $x$ such that $dist(c,x)=3$ and $parent(x)=c'$. Private edges of $c'$ can be one of the three types as follows:}\\
				\begin{enumerate}
					\item[(a)] At least one distance-$1$ private edge of $c'$ with respect to $S$ is not incident to $x$: \emph{Proof is similar to the one we have seen in previous sub-case ($(a)$)}.
					\item[(b)] All the distance-$1$ private edges with respect to $S$ are incident to $x$ : \emph{Proof is similar to the one we have seen in previous sub-case ($(b)$)}.
					
					\item[(c)] $c'$ has no distance-$1$ private edge with respect to $S$: \emph{Proof is similar to the one we have seen in previous sub-case ($(c)$)}.
					
				\end{enumerate}
			\end{enumerate}
		\end{enumerate}
	\end{ca}
\end{proof}

Combining Theorem \ref{theo:gammave=ive} and Theorem \ref{theo:TinmathscrT}, we have the main result of this section.

\begin{theo}
For a tree $T$ with at least $3$ vertices, $\gamma_{ve}(T)=i_{ve}(T)$ if and only if $T\in \mathscr{T}$.
\end{theo}

\section{Conclusions}
\label{sec:conclu}

We proposed a linear time algorithm for ve-domination problem in block graphs and also pointed out that independent ve-domination problem can also be solved using similar technique. Further, we proved that finding minimum ve-dominating set is NP-complete for undirected path graphs. Finally, we characterized the trees for which $\gamma_{ve}=i_{ve}$. It would be interesting to study this problem in other subclasses like interval graphs, directed path graphs etc. Also characterization of graphs having equal ve-domination parameters is another interesting problem.

\bibliographystyle{alpha}
\addcontentsline{toc}{section}{Bibliography}
\bibliography{VEDom}

\end{document}